\pdfoutput=1
\RequirePackage{fix-cm}
\documentclass[reqno]{amsproc}
\usepackage{amssymb}
\usepackage{hyperref}
\usepackage{microtype}
\usepackage{lmodern}
\usepackage{booktabs}

\usepackage{tikz}
\usetikzlibrary{decorations.pathreplacing}

\usepackage{graphicx}

\usepackage[citation-order,nobysame]{amsrefs}
\usepackage{xyzbib}

\usepackage{mathtools}
\mathtoolsset{showonlyrefs}
\allowdisplaybreaks

\numberwithin{equation}{section}
\let\cite=\cites

\newtheorem{theorem}{Theorem}
\newtheorem{proposition}{Proposition}
\newtheorem{conjecture}{Conjecture}
\newtheorem{lemma}{Lemma}
\DeclareMathOperator*{\res}{\mathrm{res}}

\DeclareMathOperator*{\ad}{\mathrm{ad}}
\DeclareMathOperator*{\aad}{\mathrm{aad}}

\let\leq=\leqslant
\let\geq=\geqslant

\newcommand{\ket}[1]{\left\lvert #1 \right\rangle}

\newcommand{\rmS}{\mathrm{S}}

\begin{document}


\title{Symmetries of the periodic Fredkin chain}

\author{Andrei G. Pronko}
\address{Steklov Mathematical Institute, 
Fontanka 27, 191023 Saint Petersburg, Russia}
\email{a.g.pronko@gmail.com}

\begin{abstract}  
The Fredkin chain 
is a spin-$1/2$ model with interaction of three nearest neighbors. 
In the case of periodic boundary conditions, the ground state is degenerate 
and can be described in terms of equivalence classes of Dyck paths. 
We introduce two operators commuting with the Hamiltonian  
which play the roles of lowering and raising operators when acting 
on the ground states. These operators generate the $B$- or $C$-type  
Lie algebras, depending on whether the number of 
sites $N$ is odd or even, respectively, 
with rank $n=\lceil N/2\rceil$. 
The third component of the total spin operator 
can be represented as a sum of the Cartan subalgebra elements
and some central element. In the $C$-type Lie algebra case 
(even number of sites), 
this representation coincides with a similar formula previously 
conjectured for spin-$1$ operators, 
in the context of the periodic Motzkin chain.  

\end{abstract}

\maketitle
\section{Introduction}

The Fredkin gate was originally introduced in the context 
of reversible quantum computation
\cite{FT-82}. It is also known as the controlled-SWAP gate and involves 
three qubits in its operation. Relatively recently, it was experimentally 
realized with the help of entangled photons \cite{PHFRP-16}. 

The Fredkin chain is a model of interacting   
$s=1/2$ spins, generalizing in a certain sense the 
famous Heisenberg XXX chain \cite{SK-17,SK-18}. 
In this model, the spin-$0$ component of two $s=1/2$ 
spins interacts with the third neighbor (left or right). 
Under suitably chosen open boundary 
conditions, the ground state of the Hamiltonian can be given 
as a homogeneous sum of Dyck paths. This state is highly entangled:
it shows a logarithmic growth of entanglement entropy 
with respect to the size of a subsystem. A similar phenomenon 
is known in the context of the Motzkin chain, 
which is model of nearest-neighbor interacting $s=1$ spins, 
and with the ground state 
given by a sum of Motzkin paths \cite{BCMNS-12,M-17}. 

Furthermore, the Fredkin chain appears to be even more related 
to the Motzkin chain, 
as it can be viewed as a half-integer spin analogue of the latter. 
Both models exhibit criticality without 
``frustration'': the energy gap vanishes for large system sizes, and 
the unique ground states minimize all individual terms of the Hamiltonians.
This property ensures stability of the ground state against 
inclusion in the Hamiltonian of term-dependent interaction parameters. 
The similarities extend to arbitrary half-integer and integer spin
versions of these models, 
referred to as ``colored'' Fredkin and Motzkin chains, respectively
\cite{MS-16,SUZKK-17,ZAK-17}. 
Further generalizations include the construction of various 
systems such as tensor networks \cite{AAZK-19} and a solvable cellular automaton 
\cite{SVG-23}. 
Specifically, the Fredkin chain and the Dyck paths 
description of its ground state   
can be connected to complex multi-matrix models 
\cite{WZYK-24} and Moore--Read quantum Hall states \cite{VPKGRP-24}. 
For results on the study of correlations, dynamics, and entanglement
in Fredkin and Motzkin spin chains, see, e.g., \cite{DaBT-19,CBG-24,MGM-24} 
and references therein. Recently, it was shown \cite{CSGV-25} 
that both models possess a pseudo-local conserved charge, and 
transport in these systems is subdiffusive.  

An interesting question to raise about the Fredkin chain is 
whether it is a quantum integrable model. 
Integrability of a spin chain 
implies that there exist Yang--Baxter structures 
that guarantee the existence of a hierarchy of 
conserved quantum integrals of motion  \cite{TF-79,S-82,TTF-83}.  
A paradigmatic example of such a system is 
the Heisenberg XXX chain; the Fredkin chain can be seen as its generalization. 
For a review of the methods exploiting 
integrability, see, e.g., the monograph \cite{KBI-93} and 
lecture notes \cite{F-96}; a contemporary exposition 
can be found in \cite{S-20}. 
Integrability has been found for the 
``free'' version of the Motzkin chain, in which   
one of the terms describing the interaction of spins is removed 
from the Hamiltonian \cite{HSK-23}.
Recently, it was observed, albeit for systems of small sizes, 
that the Motzkin chain with periodic boundary
conditions possesses a quite rich symmetry algebra \cite{P-25}, 
that could be a signal of its quantum integrability. 

The purpose of the present study is to obtain similar results 
for the periodic Fredkin chain.  
As for the Motzkin chain, our study here 
is mainly conjectural, although the central result about 
the properties of the raising and lowering operators 
is rigorous (see Theorem~\ref{Th2} below).  
Other results are collected in four conjectures,  
formulated by studying systems of small size, namely,
the chains with $N\leq 10$ sites. 

The main object we study is the set of two
operators which play the roles of lowering and raising 
operators when acting on the ground states. These operators 
decrease and increase, 
respectively, the eigenvalue of the third component of the
total spin operator, which labels these states. 
In the route of our study, we have also encountered an 
alternative, though less explicit, representation for these 
operators, in which they are expressed in terms of other 
components of the total spin operator. This immediately implies that 
the same operators also commute with the  Hamiltonian of the 
Heisenberg XXX spin-$1/2$ chain.   
We conjecture that these operators generate the $B$- or $C$-type  
Lie algebras, depending on whether the number of 
sites $N$ is odd or even, respectively, 
with rank $n=\lceil N/2\rceil$. Furthermore, there exists 
a central element of the symmetry algebra, which can be given as a sum of 
the third component of the total spin operator 
and Cartan subalgebra elements. Surprisingly, 
we find that in the $C$-type case (even number of sites) 
this representation coincides with a similar formula that previously 
appeared in the context of the periodic Motzkin chain. 
This observation suggests  
that the connection between the third component of the total spin operator, 
Cartan subalgebra elements, and the central element is 
independent of the $\mathfrak{sl}_2$ algebra representation.  

The paper is organized as follows. In the next section, 
we review known basic results about the periodic 
Fredkin chain. In Section 3 we introduce the raising and lowering 
operators and prove the main result about their properties. 
In Section 4 we study the algebra they generate and show that 
the third component of the total spin operator contains a central 
element extending this algebra. 
We conclude in Section 5 with a brief discussion of the results and 
possible proofs of the formulated conjectures.   

\section{The Hamiltonian, Dyck paths, and the ground states}

In this section, we recall the main ingredients from \cite{SK-17} 
about the Fredkin chain, with a focus on periodic boundary conditions. 

\subsection{The Hamiltonian}
 
We begin with fixing the notation. We denote  
basis vectors in $\mathbb{C}^2$ by
\begin{equation}
\ket{\uparrow}=
\begin{pmatrix}
1 \\ 0 
\end{pmatrix},
\qquad 
\ket{\downarrow}=
\begin{pmatrix}
0 \\ 1
\end{pmatrix}.
\end{equation}
The basis in $\mathrm{End}(\mathbb{C}^2)$ is provided 
by the Pauli matrices
\begin{equation}\label{spin1rep}
\sigma^{+}=
\begin{pmatrix}
0 & 1 \\  0 & 0  
\end{pmatrix},\qquad
\sigma^{-}=
\begin{pmatrix}
0 &  0 \\ 1 & 0 
\end{pmatrix},\qquad
\sigma^{z}=
\begin{pmatrix}
1 & 0 \\ 0 & -1  
\end{pmatrix}.
\end{equation}
We will also use projectors onto the spin-up and spin-down states:
\begin{equation}
n^\uparrow=
\frac{1}{2}(1+\sigma^z),\qquad
n^\downarrow=
\frac{1}{2}(1-\sigma^z).
\end{equation} 
The space of states of the $N$-site Fredkin chain is 
the vector space $(\mathbb{C}^2)^{\otimes N}$, in which 
the $j$th factor corresponds to the $j$th site of the spin chain. 
The basis vectors of $(\mathbb{C}^2)^{\otimes N}$ are denoted by
\begin{equation}\label{BasisKets}
\ket{\ell_1\ell_2\dots\ell_N}=\ket{\ell_1}\otimes \ket{\ell_2}
\otimes\dots\otimes \ket{\ell_N}, \qquad \ell_1,\ell_2,\dots,\ell_N=
\mathrm{\uparrow,\downarrow},
\end{equation}
and by $\sigma_j^{\pm,z}$ --- operators acting as Pauli matrices 
in the $j$th copy of $(\mathbb{C}^2)^{\otimes N}$. For later 
reference, we need the ``total spin operator'', for which we 
choose a different normalization of its components, 
\begin{equation}\label{Stotal}
\mathrm{S}^\pm=\sum_{j=1}^N\sigma_j^\pm,
\qquad
\rmS^z= \frac{1}{2}\sum_{j=1}^N \sigma_j^z,
\end{equation}   
such that the eigenvalues of $\rmS^z$ are
$S^z=-\frac{N}{2},-\frac{N}{2}+1,
\dots,\frac{N}{2}-1,\frac{N}{2}$ 
(we follow the convention of using upright 
capital letters for ``global'' operators and  
their italic variants for their eigenvalues).

Consider two adjacent copies of $\mathbb{C}^2$, say, the $j$th and $(j+1)$th.
Important objects are the permutation operator $\mathrm{P}_{j,j+1}$,
which exchanges these copies, 
and the projector $\Pi_{j,j+1}$ onto the spin-$0$ state, 
which is essentially the antisymmetrizer, 
\begin{equation}
\Pi_{j,j+1}=\frac{1}{2}(1-\mathrm{P}_{j,j+1}).
\end{equation}
The Hamiltonian density of the  
Fredkin spin chain 
describes an interaction of the spin-$0$ component of two $s=1/2$ spins with 
a neighboring third such spin, as follows: 
\begin{equation}\label{Fredkin}
\mathrm{F}_{j,j+1,j+2}=n_j^\uparrow\, \Pi_{j+1,j+2}
+\Pi_{j,j+1}\, n^\downarrow_{j+2}.
\end{equation}
The periodic boundary conditions mean that the Hamiltonian is the 
cyclic sum,
\begin{equation}\label{Hpbc}
\mathrm{H}=\sum_{j=1}^N
\mathrm{F}_{j,j+1,j+2},
\end{equation}
where identification of sites 
by modulus $N$ is assumed. 

The operator
\begin{equation}\label{Cop}
\mathrm{C}=\mathrm{P}_{1,2} \cdots \mathrm{P}_{N-2,N-1} \mathrm{P}_{N-1,N}
\end{equation}
is the cyclic shift operator $\mathrm{C}$, acting on sites as 
$i\mapsto i+1$, and it is an integral of motion, 
\begin{equation}
\left[\mathrm{C},\mathrm{H}\right]=0.
\end{equation}
The operator $\rmS^z$ is also an integral of motion,
\begin{equation}
\left[\rmS^z,\mathrm{H}\right]=0.
\end{equation}
Thus, eigenvalues of both the cyclic shift operator $\mathrm{C}$ 
and the third component of the total spin operator $\rmS^z$
can be used to label eigenstates of the Hamiltonian.

\subsection{Dyck paths}

We only summarize the construction developed in \cite{SK-17} 
to the extent that is sufficient to describe ground 
states of the Hamiltonian \eqref{Hpbc} below.

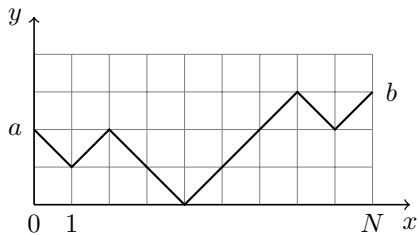
\begin{figure}
\centering

\begin{tikzpicture}[scale=.5]

\draw[help lines] (0,0) grid (9,4);


\draw [thick] (0,2)--(1,1)--(2,2)--(4,0)--(5,1)--(6,2)--(7,3)--(8,2)--(9,3);

\draw [semithick] [->] (0,0)--(10,0);
\draw [semithick] [->] (0,0)--(0,5);

\node at (0,-.5) {$0$};
\node at (10,-.5) {$x$};
\node at (-.5,5) {$y$};
\node at (1,-.5) {$1$};
\node at (9,-.5) {$N$};
\node at (-.5,2) {$a$};
\node at (9.5,3) {$b$};

\end{tikzpicture}	

\caption{A Dyck path belonging to the class $C_{a,b}(N)$, with  
$a=2$, $b=3$, and $N=9$}
\label{fig-DyckPath}
\end{figure}

Under a Dyck path we will understand a random walk on the square lattice 
with vertices labeled by $(x,y)$ where 
$x$ and $y$ are non-negative integers. 
A Dyck path of length $N$  starts
at $(x,y)=(a,0)$ and ends at $(x,y)=(b,N)$, 
at each step $\Delta x =1$ and $\Delta y= \pm1$, with the condition that 
at least once the path hits the $x$-axis, i.e.,
$\min y =0$ along the path. Note that $a$, $b$, and $N$ are related: 
there must exist $k\geq 0$ such that $N=2k+a+b$. 
Following \cite{SK-17} we will call all such paths 
as belonging to the equivalence class $C_{a,b}(N)$. 
An example is shown in \figurename~\ref{fig-DyckPath}.

The paths in the class $C_{a,b}(N)$ are equivalent up to 
the recursive procedure where at each iteration 
a pair of consecutive up and down steps, forming a peak, can be moved anywhere
along the path (in \cite{SK-17} these iterations were called Fredkin moves). 
Thus, one can transform 
a given Dyck path to some ``reference'' path which 
can be chosen, for example, as the path having 
first $a$ steps down, next $k$ pairs of peaks, and, finally, $b$ steps up. 
 
Apparently, all possible Dyck paths of length $N$ are 
in one-to-one correspondence with the basis vectors 
of $(\mathbb{C}^2)^{\otimes N}$, the state space of the Fredkin chain. 
The correspondence is based on a simple observation that 
for each letter $\ell_j$ in \eqref{BasisKets} one can associate step 
up or step down according to whether $\ell_j=\uparrow$ or $\ell_j=\downarrow$. 
Thus, a string (or, a ``word'') $\ell_1\dots\ell_N$ uniquely defines 
the Dyck path, and this can be used for its notation. 

\begin{figure}
\centering

\begin{tikzpicture}[scale=.5]
\node at (1.5,-1.5) {$C_{0,3}(3)$};
\draw [thick] (0,0)--(3,3);
\draw [help lines] (0,0) grid (3,3);
\node at (9.5,-1.5) {$C_{0,1}(3)$};
\draw [thick] (6,0)--(7,1)--(8,0)--(9,1); 
\draw [help lines] (6,0) grid (9,1);
\draw [thick] (10,0)--(12,2)--(13,1); 
\draw [help lines] (10,0) grid (13,2);
\node at (17.5,-1.5) {$C_{1,2}(3)$};
\draw [thick] (16,1)--(17,0)--(19,2); 
\draw [help lines] (16,0) grid (19,2);
\end{tikzpicture}	

\caption{Equivalence classes of Dyck paths
in the case $N=3$; the remaining can be obtained by 
a mirror transformation}
\label{fig-ThreeSites}
\end{figure}
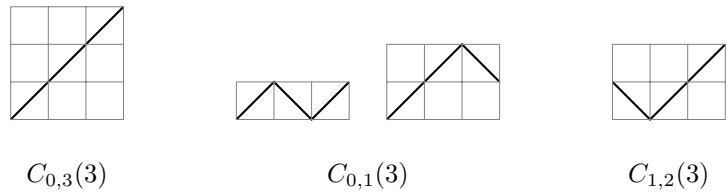

Division of Dyck paths into equivalence classes introduces certain 
order among the basis vectors. In particular, vectors associated with 
the class $C_{a,b}(N)$ are eigenvectors of $\rmS^z$ with the eigenvalue  
$S^z=\frac{1}{2}(b-a)$. Furthermore, one can introduce 
states $\ket{C_{a,b}(N)}$ 
as homogeneous sums over Dyck paths belonging to 
$C_{a,b}(N)$, namely,  
\begin{equation}
\ket{C_{a,b}(N)}=
\sum_{\ell_1\ell_2\dots\ell_N\in C_{a,b}(N)}
\ket{\ell_1\ell_2\dots\ell_N}. 
\end{equation}
For example, in the case $N=3$ the 
allowed Dyck paths, see \figurename~\ref{fig-ThreeSites}, 
yield  
\begin{equation}
\ket{C_{0,3}(3)}=\ket{\uparrow\uparrow\uparrow},
\qquad
\ket{C_{0,1}(3)}=\ket{\uparrow\downarrow\uparrow}
+\ket{\uparrow\uparrow\downarrow},
\qquad
\ket{C_{1,2}(3)}=\ket{\downarrow\uparrow\uparrow},
\end{equation}
and there are also the states $\ket{C_{3,0}(3)}$, $\ket{C_{1,0}(3)}$, and 
$\ket{C_{2,1}(3)}$ obtained by reversal of spins of these states
(mirror transformation of paths), respectively.  

The Dyck paths for $N$ even belonging to the class 
$C_{0,0}(N)$ correspond to the Dyck paths in their standard definition, i.e., 
these are paths that start at $(x,y)=(0,0)$ and end at 
$(x,y)=(0,N)$ with the condition that they are not allowed to 
pass below the $x$-axis. The corresponding state 
$\ket{C_{0,0}(N)}$ is the unique ground state of 
the open Fredkin chain introduced in 
\cite{SK-17,SK-18}.

\subsection{Ground states}

In the case of periodic boundary conditions, 
the ground states corresponding to 
zero eigenvalues of the Hamiltonian can be constructed as 
a superposition of states $\ket{C_{a,b}(N)}$ with 
$a$ and $b$  related to an eigenvalue of the operator $\rmS^z$.
A key observation leading to expressions for the ground states 
is an extension of 
the Fredkin moves by periodicity, see \cite{SK-17,SK-18}, for details.
To emphasize that this is an established result, we formulate it here as 
a theorem. 

\begin{theorem}[Salberger, Korepin \cite{SK-17,SK-18}] \label{Th1}
For each value of $S^z\in \{-\frac{N}{2},-\frac{N}{2}+1,\dots,
\frac{N}{2}-1,\frac{N}{2}\}$ 
there exists one and only one ground state $\ket{v_{S^z}}$ 
of the periodic Fredkin chain Hamiltonian \eqref{Hpbc} with 
the zero eigenvalue, 
which corresponds to the eigenvalue $C=1$ of the 
cyclic shift operator, of the form 
\begin{equation}\label{vecSz}
\ket{v_{S^z}}=
\sum_{\substack{a,b\geq 0\\ b-a =2S^z}}\ket{C_{a,b}(N)}.
\end{equation}
In the case of $N$ even, $N=2n$, and $S^z=0$ there exists 
one more ground state which corresponds to the eigenvalue $C=-1$, 
of the form
\begin{equation}\label{v0minus}
\ket{v_0^{-}}= \sum_{k=0}^{n}(-1)^k \ket{C_{k,k}(2n)}.
\end{equation} 
\end{theorem}

Thus\footnote{In \cite{SK-17,SK-18} the degeneracies for odd and even $N$ 
were misprinted as $N$ and $N+1$, respectively.}, 
in the case of $N$ odd 
the degeneracy of the ground state 
is $N+1$ and in the case of $N$ even, it is $N+2$.  

\begin{figure}
\centering

\begin{tikzpicture}[scale=.5]
\node at (4.5,-1.5) {$C_{0,0}(4)$};
\draw [thick] (0,0)--(1,1)--(2,0)--(3,1)--(4,0); 
\draw [help lines] (0,0) grid (4,1);
\draw [thick] (5,0)--(7,2)--(9,0);
\draw [help lines] (5,0) grid (9,2);
\node at (14,-1.5) {$C_{2,2}(4)$};
\draw [thick] (12,2)--(14,0)--(16,2); 
\draw [help lines] (12,0) grid (16,2);
\end{tikzpicture}	
\vspace{.2in}\par
\begin{tikzpicture}[scale=.5]
\node at (7,-1.5) {$C_{1,1}(4)$};
\draw [thick] (0,1)--(1,0)--(2,1)--(3,0)--(4,1);
\draw [help lines] (0,0) grid (4,1);
\draw [thick] (5,1)--(6,0)--(8,2)--(9,1);
\draw [help lines] (5,0) grid (9,2);
\draw [thick] (10,1)--(11,2)--(13,0)--(14,1);
\draw [help lines] (10,0) grid (14,2);
\end{tikzpicture}	

\caption{Three classes $C_{a,a}(N)$ of the Dyck paths for $N=4$; 
the first line (even $a$) gives rise to the state $\ket{v_0^\text{even}}$, and 
the second line (odd $a$) to the state $\ket{v_0^\text{odd}}$}
\label{fig-FourSites}
\end{figure}
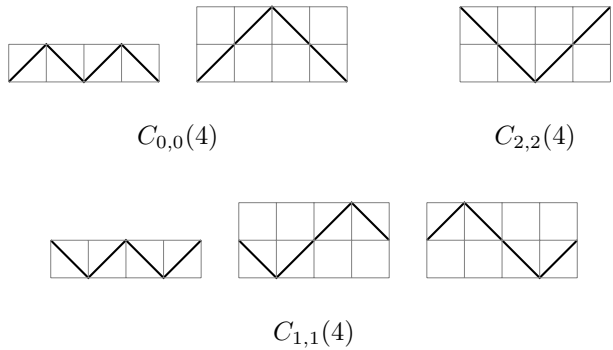

Appearance of two ground states for $N$ even and $S^z=0$ 
comes from the fact that the considerations involving the periodic 
Fredkin moves lead to existence of two eigenstates, with even and odd 
values of $a$ for the classes $C_{a,a}(N)$. 
An example of $N=4$ is shown in \figurename~\ref{fig-FourSites}.
These states are
\begin{equation}\label{even-odd}
\ket{v_0^\text{even}}=\sum_{k=0}^{n}\ket{C_{2k,2k}(2n)},\qquad 
\ket{v_0^\text{odd}}=\sum_{k=0}^{n-1}\ket{C_{2k+1,2k+1}(2n)},
\end{equation}
and they can be mapped to each other by the cyclic shift operator,
\begin{equation}\label{Cvv}
\mathrm{C}\ket{v_0^\text{even}}=\ket{v_0^\text{odd}},\qquad 
\mathrm{C}\ket{v_0^\text{odd}}=\ket{v_0^\text{even}}.
\end{equation}
Their linear combinations 
$\ket{v_0^\pm}=\ket{v_0^\text{even}}\pm\ket{v_0^\text{odd}}$ are 
eigenstates of the cyclic shift operator with the eigenvalues $C=\pm1$, thus leading 
to the states $\ket{v_0}\equiv \ket{v_0^{+}}$ and $\ket{v_0^{-}}$ 
mentioned above.

\section{Lowering and raising operators}

Now we turn to the results. We have  
obtained them mainly by a direct study of the 
model with $N\leq 10$ sites. The central result, about explicit 
expressions for raising and lowering operators, is given here as a theorem. 
The remaining results (in this and the next section) are summarized in four
conjectures. 

\subsection{Symmetry generators}

In \cite{SK-17,SK-18} it was indicated that the 
states $\ket{v_{S^z}}$ are nothing but the ground states of the 
spin-$1/2$ ferromagnetic Heisenberg XXX chain (see, e.g., \cite{F-96}). 
As such, they 
can be obtained from each other by acting with the 
operators $\rmS^{+}$ and $\rmS^{-}$, 
which increase and decrease, respectively, the 
value of $S^z$. However, these operators have no similar role in the context
of the Fredkin chain since they do not commute with 
the Hamiltonian \eqref{Hpbc}. 

One of our main results here is that there exists another pair of operators 
which play a similar role of 
raising and lowering operators when acting on the states $\ket{v_{S^z}}$, 
but commuting with the Hamiltonian \eqref{Hpbc}.  

\begin{theorem}\label{Th2}
The operators  
\begin{equation}\label{SigmapmSum}
\Sigma^\pm=
\sum_{\substack{r_1,\dots,r_N\in \{-1,0,1\}\\ r_1+\dots+r_N=\pm 1}} 
\sigma_1^{r_1}\cdots \sigma_N^{r_N},
\end{equation}
where $\sigma_i^{0}= 1$ and $\sigma_i^{\pm 1}= \sigma_i^{\pm}$, 
or, equivalently,
\begin{equation}\label{SigmapmRes}
\Sigma^\pm=\res_{\lambda=0}\,\prod_{i=1}^N 
\Bigl(\lambda^{-1} \sigma_i^{\pm} 
+1 +\lambda\sigma_i^{\mp}\Bigr),
\end{equation}  
when acting on the ground states $\ket{v_{S^z}}$ 
play the role of raising and lowering operators, respectively,
\begin{equation}\label{Sigmavec}
\Sigma^\pm\ket{v_{S^z}}
=
\begin{cases}
c_\pm(S^z)\ket{v_{S^z\pm1}}, & S^z\ne \pm N/2,
\\
0, & S^z=\pm N/2,
\end{cases}
\end{equation}
where $c_\pm(S^z)\ne 0$ are some constants. These operators  
commute with the Hamiltonian,
\begin{equation}\label{SpmH}
\left[\Sigma^\pm,\mathrm{H}\right]=0.
\end{equation}
\end{theorem}
We give a proof below at the end of this section. Surprisingly, the 
assertion of the theorem about commutativity with the Hamiltonian
can be seen as a corollary of the  
technically stronger statement (see Proposition~\ref{Prop1} below).

The operators $\Sigma^\pm$ 
annihilate the state $\ket{v_0^-}$ 
specific for $N$ even,
\begin{equation}\label{Svminus=0}
\Sigma^\pm\ket{v_0^-}=0.
\end{equation} 
This property can be seen as a particular case of the following 
more general observation.
\begin{conjecture}
The operators $\Sigma^\pm$ annihilate all non-cyclic invariant ($C\ne 1$)
eigenstates of the Hamiltonian.    
\end{conjecture}

\subsection{A relation to the total spin operator}

As it is clear from \eqref{SigmapmSum}, the matrices
representing operators $\Sigma^\pm$ 
are built of $0$'s and $1$'s; 
these matrices are nilpotent, $(\Sigma^\pm)^{N+1}=0$, though not 
strictly upper or lower triangular ones.   
The operators $\Sigma^\pm$ 
are non-local, contrary to the operators $\rmS^\pm$,
and at a first glance they are unrelated. However, it turns out that   
$\Sigma^\pm$'s can be expressed in terms of $\rmS^{\pm}$'s. 

\begin{conjecture}
For the operators $\Sigma^\pm$ there exists the representation
\begin{equation}
\Sigma^\pm =\sum_{k=1}^{\lceil N/2\rceil }
\gamma_k \left(\aad \rmS^\pm \aad \rmS^\mp\right)^{k-1}
\rmS^\pm,
\end{equation}
where $\gamma_k$ are some coefficients and $\aad$ denotes the 
anti-adjoint action, $(\aad a)\, b \equiv \{a, b\}=ab+ba$.
\end{conjecture}
For example, in the case $N=3$ we find
\begin{equation}
\Sigma^\pm = -\frac{1}{4}\rmS^\pm+\frac{1}{8}\{\rmS^\pm,\{\rmS^\mp,\rmS^\pm\}\}.
\end{equation}
Coefficients $\gamma_k$ for cases $N\leq 10$ are
given in Table~\ref{gks}. 

\begin{table}[!ht]
\centering
\caption{Coefficients $\gamma_k$}
\label{gks}
\begin{tabular}{r|ccccc}
\toprule 
$N$ & $\gamma_1$ & $\gamma_2$ & $\gamma_3$ & $\gamma_4$ & $\gamma_5$ \\
\midrule 
3& $-\frac{1}{4}$  & $\frac{1}{8}$ \\[1ex]
4& $-\frac{3}{4}$  & $\frac{1}{8}$ \\[1ex]
5& $\frac{1}{8}$   & $-\frac{7}{96}$ & $\frac{1}{192}$\\[1ex]
6& $\frac{5}{8}$   & $-\frac{13}{96}$ & $\frac{1}{192}$\\[1ex]
7
& $-\frac{5}{64}$ & $\frac{37}{768}$ 
& $-\frac{11}{2304}$ & $\frac{1}{9216}$ \\[1ex]
8 
& $-\frac{35}{64}$ & $\frac{101}{768}$ 
& $-\frac{17}{2304}$ & $\frac{1}{9216}$ \\[1ex]  
9
& $\frac{7}{128}$ & $-\frac{533}{15\,360}$ & $\frac{727}{184\,320}$ 
& $-\frac{5}{36\,864}$ & $\frac{1}{737\,280}$ \\[1ex]
10
& $\frac{63}{128}$ & $-\frac{641}{5120}$ & $\frac{509}{61\,440}$ 
& $-\frac{7}{36\,864}$ & $\frac{1}{737\,280}$ \\
\bottomrule 
\end{tabular}
\end{table}

\subsection{Proof of Theorem~\ref{Th2}}

To show that \eqref{Sigmavec} holds, we start with rewriting 
one of the results of
Theorem~\ref{Th1}, namely, \eqref{vecSz} in a form suitable for application 
of formula \eqref{SigmapmSum} for the operators $\Sigma^\pm$.  
\begin{lemma}
The cyclic-invariant ($C=1$) ground states $\ket{v_{S^z}}$ 
can be written as follows:
\begin{equation}\label{vecSum}
\ket{v_{N/2-m}}=\sum_{1\leq k_1<k_2<\dots<k_m\leq N}
\sigma^{-}_{k_1}\sigma^{-}_{k_2}\cdots\sigma^{-}_{k_m} \ket{\Uparrow}.
\end{equation}
Equivalently,  
\begin{equation}\label{vecRes}
\ket{v_{N/2-m}}=\res_{\lambda=0} \lambda^{m-1}
\prod_{i=1}^{N}\left(\lambda^{-1}\sigma^{-}_{i}+1\right) \ket{\Uparrow}
\end{equation} 
or 
\begin{equation}\label{vecRes2}
\ket{v_{N/2-m}}=\res_{\lambda=0} \lambda^{-m-1}
\prod_{i=1}^{N}\left(\lambda\sigma^{-}_{i}+1\right) \ket{\Uparrow}.
\end{equation} 
Here, $\ket{\Uparrow}$ is the state ``all spins up'', 
$\ket{\Uparrow}=\ket{\uparrow}^{\otimes N}$.
\end{lemma}
Formula \eqref{vecSum} is simply \eqref{vecSz} written 
in terms of the Pauli spin operators; \eqref{vecRes} and \eqref{vecRes2} 
are analogues of \eqref{SigmapmRes}. 

Now, one can prove the result of 
action of the 
operators $\Sigma^\pm$ 
on the vectors $\ket{v_{S^z}}$. Using the residue formulas     
\eqref{SigmapmRes} and \eqref{vecRes}, we find
\begin{align}
\Sigma^\pm\ket{v_{N/2-m}}
&
=\res_{\lambda=0}\res_{\mu=0}
\lambda^{m-1}\prod_{i=1}^N 
\left(\mu^{-1}\sigma_i^{\pm}+1+\mu\sigma_i^{\mp}\right)
\left(\lambda^{-1}\sigma_i^{-}+1\right) \ket{\Uparrow}
\\ & 
=\res_{\lambda=0}\res_{\mu=0}\lambda^{m-1}\prod_{i=1}^N 
\left((\mu^{\pm 1}+\lambda^{-1})\sigma_i^{-}+1+\mu^{\mp1}\lambda^{-1}
\sigma_i^{+}\sigma_i^{-}\right)
\ket{\Uparrow}
\\ & 
=\res_{\lambda=0}\res_{\mu=0} 
\lambda^{m-1}\left(1+\mu^{\mp1}\lambda^{-1}\right)^N
\prod_{i=1}^N 
\left(\mu^{\pm 1}\sigma_i^{-}+1\right) \ket{\Uparrow}
\\ & 
=\binom{N}{m}
\res_{\mu=0} \mu^{\pm m} 
\prod_{i=1}^N 
\left(\mu^{\pm 1}\sigma_i^{-}+1\right) \ket{\Uparrow},
\end{align}
and \eqref{Sigmavec} follows, after taking into account 
\eqref{vecRes} (for $\Sigma^{-}$) or  
\eqref{vecRes2} (for $\Sigma^{+}$).

Let us now consider commutativity of the operators $\Sigma^\pm$ with the 
Hamiltonian, relation \eqref{SpmH}. It is useful to consider more
general operators:
\begin{equation}\label{SigmaQ}
\Sigma^Q_{i_1,\dots,i_n}=
\sum_{\substack{r_1,\dots,r_n\in \{-1,0,1\}\\ r_1+\dots+r_n=Q}} 
\sigma_{i_1}^{r_1}\cdots \sigma_{i_n}^{r_n},
\qquad 
Q=0,\pm 1,\dots, \pm n.
\end{equation}
Here, as above, $\sigma_i^{0}\equiv 1$, $\sigma_i^{\pm 1}\equiv 
\sigma_i^{\pm}$, and 
the subscripts $i_1,\dots,i_n$, indicate copies of $\mathbb{C}^2$ where the
Pauli spin  operators act, $1\leq i_1<\dots<i_n\leq N$. 
The superscript $Q$ can be interpreted as a ``charge'' of the operator. Note that 
with this notation $\Sigma^\pm\equiv \Sigma^{\pm 1}_{1,\dots,N}$, and below 
we drop the subscripts when $n=N$, i.e., writing 
$\Sigma^Q\equiv \Sigma^Q_{1,\dots,N}$. The following observation is crucial.  
\begin{proposition}\label{Prop1}
The Hamiltonian density of the Fredkin chain \eqref{Fredkin}
satisfies
\begin{equation}\label{CMSF}
[\Sigma^Q,\mathrm{F}_{i,i+1,i+2}]=0,\qquad Q=0,\pm1,\dots,\pm N.
\end{equation}
\end{proposition}
\begin{proof}

Because of the cyclic invariance of the operators $\Sigma^Q$, 
to prove \eqref{CMSF} 
it is sufficient to consider the case $i=1$, i.e., to show that it holds for 
the operator $\mathrm{F}_{123}$. 
The operators $\Sigma^Q$ admit the decomposition 
\begin{equation}\label{SumSigma}
\Sigma^Q=\sum_{\tilde Q=-3}^{3} \Sigma_{123}^{\tilde Q} 
\Sigma_{4,\dots,N}^{Q-\tilde Q}.
\end{equation}  
Note that this decomposition simply follows from the core definition 
\eqref{SigmaQ} and it can also be derived formally by writing a 
residue formula by analogue with \eqref{SigmapmRes}, and next evaluating 
the residues after the summation. 
The decomposition \eqref{SumSigma} implies that 
\eqref{CMSF} will follow if $[\Sigma_{123}^Q,\mathrm{F}_{123}]=0$ for 
$Q=0,\pm1,\pm2,\pm3$. Explicitly, the operators appearing here are 
\begin{align}
\Sigma^{0}_{123}&=1+\sigma_1^{+}\sigma_2^{-}+\sigma_1^{-}\sigma_2^{+}
+\sigma_1^{+}\sigma_3^{-}+\sigma_1^{-}\sigma_3^{+}+\sigma_2^{+}\sigma_3^{-}
+\sigma_2^{-}\sigma_3^{+},
\\
\Sigma^{\pm 1}_{123}&=\sigma_1^{\pm} +\sigma_2^{\pm}+\sigma_3^{\pm}
+\sigma_1^{\pm}\sigma_2^{\pm}\sigma_3^{\mp}+
\sigma_1^{\pm}\sigma_2^{\mp}\sigma_3^{\pm}+
\sigma_1^{\mp}\sigma_2^{\pm}\sigma_3^{\pm},
\\
\Sigma^{\pm 2}_{123}&=\sigma_1^{\pm}\sigma_2^{\pm}
+\sigma_1^{\pm}\sigma_3^{\pm}+
\sigma_2^{\pm}\sigma_3^{\pm},
\\
\Sigma^{\pm 3}_{123}&=\sigma_1^{\pm}\sigma_2^{\pm}\sigma_3^{\pm}.
\end{align}
A direct check shows that $\mathrm{F}_{123}$ indeed 
commutes with all of them.
\end{proof}
Thus, the commutativity property \eqref{SpmH} follows simply 
from a particular case of \eqref{CMSF} at $Q=\pm1$. 
This completes the proof of Theorem~\ref{Th2}.

\section{Symmetry algebra}

Here, we focus on further discussion of properties of the operators 
$\Sigma^\pm$, such as which algebra they span and what
the role of the operator $\rmS^z$ is in this construction. 
We start with comments about one more operator 
which exists in the case of $N$ even and next proceed to the results. 

\subsection{Preliminaries}

In the case of even $N$, one can introduce one more operator, $\Xi$,  
which, when acting on the ground states 
maps the states $\ket{v_0}\equiv \ket{v_0^{+}}$ and $\ket{v_0^{-}}$
to each other, 
\begin{equation}
\Xi\ket{v_0^\pm}=\ket{v_0^\mp},
\end{equation}
and annihilates the states $\ket{v_{S^z}}$ with $S^z\ne 0$,
\begin{equation}
\Xi\ket{v_{S^z}}=0,\qquad S^z=\pm1, \dots, \pm N/2.
\end{equation}
Inspecting the structure of the states \eqref{v0minus}, it is not difficult 
to figure out that such an operator can be written in the form 
\begin{equation}\label{Xi}
\Xi=\sum_{k=0}^{N/2} (-1)^k 
\sum_{\ell_1\ell_2\dots\ell_N\in C_{k,k}(N)}
n_1^{\ell_1} n_2^{\ell_2}\dots n_N^{\ell_N},
\end{equation}
where $n_j^{\ell_j}$, $\ell_j=\uparrow,\downarrow$, are 
the spin-up and spin-down 
projection operators, and the inner summation is performed 
over the Dyck paths from the class $C_{k,k}(N)$.

As it is clear from \eqref{Xi}, the operator $\Xi$ 
is represented by a diagonal matrix with $0$'s, $1$'s and $-1$'s 
in its entries, with $0$'s responsible for cutting out all sectors 
with $S^z=\pm 1,\dots,\pm N/2$. In the sector with $S^z=0$
the entries are all nonzero, represented by 
$1$'s and $-1$'s in an equal number. In other words, 
$\Xi$ acts nontrivially only in the $S^z=0$ sector of 
the whole Hilbert space of the model. It commutes with the Hamiltonian
and anti-commutes with the cyclic shift operator: 
\begin{equation}
[\Xi,\mathrm{H}]=0,\qquad
\mathrm{C}\Xi \mathrm{C}^{-1}=-\Xi.
\end{equation}
The first relation may require a proof though we find it 
satisfied in all checks; the second one is obvious from \eqref{Xi}.  
These relations mean that $\Xi$ maps every state
in the $S^z=0$ sector to another state in this sector 
with the same eigenvalue of $\mathrm{H}$, but alternated eigenvalue of the 
cyclic shift operator, $C\mapsto -C$. 
Existence of such an unusual operator may explain a clearly visible 
(in numerical checks) double degeneracy of the whole spectrum 
of the Hamiltonian.  It complements, for the case of 
$S^z=0$, the doubling of states 
in the $S^z\ne 0$ sectors following from the spin-reversal symmetry.  

Nevertheless, we exclude $\Xi$ from our considerations below, since 
in treating the symmetry algebra of the model only operators 
which commute with both integrals of motion, $\mathrm{H}$ and $\mathrm{C}$,
are of interest. 

\subsection{The Lie algebra of raising and lowering operators}

Now we focus on establishing the algebra 
the operators $\Sigma^\pm$ generate. 
We denote this algebra by $\mathfrak{g}$.
Our findings can be summarized as follows.

\begin{conjecture}
The rank of $\mathfrak{g}$ is $n=\lceil N/2 \rceil$
and   
\begin{equation}
\mathfrak{g}=
\begin{cases}
B_n=\mathfrak{so}_{2n+1}, & N=2n-1,
\\
C_n=\mathfrak{sp}_{2n}, & N=2n.
\end{cases}
\end{equation}
\end{conjecture}

To clarify the statement, it is useful to recall 
(see, e.g., \cite{K-02,IR-18}) that a 
complex semisimple Lie algebra has a standard definition in terms of 
Chevalley generators  
$e_i, f_i, h_i$, $i=1,\dots, n$, where $n$ is the rank of the algebra,  
which satisfy the Serre relations:  
\begin{equation}\label{Serre}
\begin{gathered}
[h_i,h_j]=0,
\qquad [e_i,f_i]=h_i,
\qquad
[e_i,f_j]=0,\quad i\ne j,
\\
[h_i,e_j]=A_{ij}e_j, \qquad [h_i,f_j]=-A_{ij}f_j,
\\ 
(\ad e_i)^{1-A_{ij}}e_j=(\ad f_i)^{1-A_{ij}} f_j=0,\quad i\ne j.
\end{gathered}
\end{equation}
Here, $\ad a \equiv [a,\cdot\,{}]$, and 
the numbers $A_{ij}$ are entries of the Cartan matrix. 
One can identify the algebra by computing the Cartan matrix
that involves construction of the Chevalley generators. 

As in \cite{P-25}, to formulate the above conjecture we 
have performed the following calculation.  
In addition to 
$\Sigma^\pm$, we define 
$\Sigma^z=[\Sigma^{+},\Sigma^{-}]$ 
and construct triples $\Lambda^{\pm,z}$, $\Phi^{\pm,z}$, etc, 
introducing them recursively, first
$\Lambda^\pm=\pm\left[\Sigma^z,\Sigma^{\pm}\right]$,
$\Lambda^z=\left[\Lambda^{+},\Lambda^{-}\right]$, 
next 
$\Phi^\pm=\pm\left[\Lambda^z,\Lambda^{\pm}\right]$,
$\Phi^z=\left[\Phi^{+},\Phi^{-}\right]$, and so on.  
An important observation which can be easily done at this stage  
is that the operators $\Sigma^z,\Lambda^z,\Phi^z,\dots$ 
form an abelian algebra, and therefore they can be considered as 
linear combinations of the Cartan subalgebra elements $h_i$. 
Hence, the rank of the algebra can be identified by how many operators 
$\Sigma^z,\Lambda^z,\Phi^z,\dots$ are linearly independent.

As the rank is found, the next step is to search 
elements $e_i$ and $f_i$ in the form
\begin{equation}\label{eifi}
e_i=\rho_i\left(\Sigma^{+}+a_i \Lambda^{+}+b_i \Phi^{+}+\dots\right),
\qquad
f_i=\rho_i\left(\Sigma^{-}+a_i \Lambda^{-}+b_i \Phi^{-}+\dots\right),
\end{equation} 
where the number of terms is equal to the rank.
Solving relations $[e_i,f_j]=0$, $i\ne j$, for the coefficients 
$a_i, b_i,\dots$ one finds a unique solution up to a permutation
of triples of Chevalley generators. 
The solution is fixed by finding $\rho_i$ from 
the relations $[h_i,e_i]=2e_i$, which correspond to the canonical 
normalization of the Cartan matrix with $A_{ii}=2$.

At the last step, the off-diagonal entries 
of the Cartan matrix are computed according to 
the relations $[h_i,e_j]=A_{ij}e_j$. 
A particular order among the 
triples of Chevalley generators should be fixed, 
such that the Cartan matrix can be identified
according to the well-known classification by Dynkin diagrams. 
In our problem,
we meet Cartan matrices for the Lie algebras of $B$- and $C$-types. 
Recall that the off-diagonal nonzero entries of the Cartan matrix,
in the case of $B$-type algebra are
\begin{equation}
A_{i+1,i}=-1,\qquad
A_{i,i+1}=-1\quad (i\ne n-1),\qquad
A_{N-1,N}=-2,
\end{equation}
and in the case of $C$-type algebra are
\begin{equation}
A_{i+1,i}=-1 \quad (i\ne n-1), \qquad 
A_{N,N-1}=-2, \qquad 
A_{i,i+1}=-1. 
\end{equation}
As a self-consistency check, the relations 
$(\ad e_i)^{1-A_{ij}}e_j=0$, $i\ne j$, are verified. 
 
\subsection{A central element}

Noting that 
$[\rmS^z,\Sigma^\pm]=\pm \Sigma^\pm$, one can expect that 
the operator $\rmS^z$ can be expressed in terms of 
the Cartan subalgebra elements. It turns out 
that such a construction involves one more operator 
which appears to be a central element with respect to the generators 
of $\mathfrak{g}$.  This can be formulated as follows.

\begin{conjecture}
There exists a nontrivial operator $p$ 
which is a central element 
with respect to the algebra $\mathfrak{g}$, and it has the form
\begin{equation}\label{Sz=psum}
p=\rmS^z-\sum_{i=1}^{\lceil N/2 \rceil}\alpha_i h_i,
\end{equation}  
where $\rmS^z$ is the third component of total spin operator, 
$\alpha_i$ are some coefficients, and 
$h_i$ are the Cartan subalgebra elements.  
\end{conjecture}

We call $p$ a central element 
since it commutes with the operators $\Sigma^\pm$ and hence with 
all the Chevalley generators, $[p,e_i]=[p,f_i]=[p,h_i]$. 

Formula \eqref{Sz=psum} can be viewed as an expression for 
the operator $\rmS^z$ in terms of elements of $\mathfrak{g}$ and 
the operator $p$. The coefficients $\alpha_k$ for $N\leq 10$ are given 
in Table~\ref{aks}. 

\begin{table}[!ht]
\centering
\caption{Coefficients $\alpha_k$}
\label{aks}
\begin{tabular}{r|ccccc}
\toprule 
$N$ & $\alpha_1$ & $\alpha_2$ & $\alpha_3$ & $\alpha_4$ & $\alpha_5$ \\
\midrule 
3 & $2$ & $\frac{3}{2}$ \\[1ex]
4 & $2$ & $\frac{3}{2}$ \\[1ex]
5 & $\frac{5}{2}$ & $4$ & $\frac{9}{2}$ \\[1ex]
6 & $3$ & $5$ & $3$ \\[1ex]
7 & $\frac{7}{2}$ & $6$ & $\frac{15}{2}$ & $8$ \\[1ex]
8 & $4$ & $7$ & $9$ & $5$ \\[1ex] 
9 & $\frac{9}{2}$ & $8$ & $\frac{21}{2}$ & $12$ & $\frac{25}{2}$\\[1ex]
10 & $5$ & $9$ & $12$ & 14 & $\frac{15}{2}$\\  
\bottomrule 
\end{tabular}
\end{table}

Coincidence of the coefficients for $N=3$ and $N=4$ can be 
explained by the appearance of the same algebra ($B_2=C_2$) in these two cases. 
Furthermore, an interesting conclusion which 
can be made by inspecting these examples,  
is that in the case of $C$-type Lie algebra ($N$ even) formula \eqref{Sz=psum}
exactly coincides with that obtained in the 
context of the periodic Motzkin chain (of length $N/2$) 
\cite{P-25}. This suggests that
relation \eqref{Sz=psum} is probably independent of the choice of the initial 
representation of the $\mathfrak{sl}_2$ algebra of local spin operators. 

Inspecting values of $\alpha_k$'s 
in Table~\ref{aks} one can make some further conjectures. For example, 
one can guess that $\alpha_1 =N/2$ for $N\geq 4$, and $\alpha_2=N-1$ 
for $N\geq 5$. It looks also plausible 
that $\alpha_3=\frac{3}{2}N-3$ for $N\geq 7$. 

However, to gain more insight into these numbers, outputs are needed for 
further values of $N$ (beyond the $N=10$ case). This is not quite 
a simple computational task, as it requires explicit construction 
of the Chevalley generators to establish the corresponding 
relation \eqref{Sz=psum}. Along the way, one has to solve the equations
$[e_i,f_j]=0$, $i\ne j$, for the unknowns $a_i, b_i,\dots$ 
in \eqref{eifi}. These equations are quadratic with integer coefficients 
that grow enormously with $N$ (e.g., in the $N=10$ case, they consist 
of almost two hundred digits).

\section{Conclusion}

In this paper, we have considered 
the Fredkin spin chain with periodic boundary conditions. 
Our starting point is the result of Salberger and Korepin 
\cite{SK-17,SK-18} about 
the explicit form of the ground states (Theorem~\ref{Th1}).  
The results are summarized in one theorem and four conjectures.
The latter have been formulated by studying the chain of length $N\leq 10$.

We have shown that there exist raising and lowering operators 
which map these states to each other and which commute 
with the Hamiltonian (Theorem~\ref{Th2}). We have also 
provided several conjectures about other properties of these operators.
Conjecture 1 says that these operators actually act non-trivially 
only in the subspace of cyclic invariant states. Conjecture 2 shows that 
these operators can be (somewhat unexpectedly) expressed in terms of  
the non-diagonal components of the total spin operator. Conjecture 3 states 
that they span in $B$- or $C$-type Lie algebra, according to the parity of 
the number of sites of the chain. Conjecture 4 asserts that the symmetry 
algebra of the model is actually wider and extended by 
a central element contained in the third component 
of the total spin operator.      

In our previous study devoted to the Motzkin chain \cite{P-25} 
the result analogous to Theorem~\ref{Th2} was formulated as a conjecture. 
In fact, the proof we give here for Theorem~\ref{Th2} can be directly 
generalized to the spin-$1$ case of Motzkin chain. This is intimately 
related to fact that the lowering and raising operators 
have essentially the same structure for both models. 

Conjecture 1 can also be formulated in the context of Motzkin chain. 
It is not however clear how it can be proven in either case.     
Concerning Conjecture 2, it seems plausible that it can be proven 
by suitably applying the technique exposed in \cite{ZF-11}. 
Conjectures 3 and 4 are direct analogues of 
the corresponding conjectures made in the context of Motzkin chain in 
\cite{P-25}. We have observed that the 
central element for the Fredkin chain with even number of sites 
appears to be related to 
the third component of total spin operator 
by exactly the same relation as was found in the context 
of the Motzkin chain.  In both cases the symmetry algebra 
is the $C$-type Lie algebra of the same rank. It would be interesting  
to understand group-theoretical 
backgrounds of these results. In addition, the coefficients 
$\gamma_i$ and $\alpha_i$ appearing in Conjecture 2 and Conjecture 4, 
respectively, certainly have some combinatorial interpretations that
can also be a subject of further study.

\section*{Acknowledgments}

The author is indebted for stimulating discussions to  
Nikolai M. Bogoliubov, Vladimir E. Korepin, and Vitaly O. Tarasov.
The author also thanks anonymous referees for useful remarks. 

\bibliography{sympfre_bib}
\end{document}